\documentclass[preprint,preprintnumbers,amsmath,amssymb]{revtex4}

\newcommand{\apar}{\Delta \alpha _{\parallel}}
\newcommand{\aper}{\Delta \alpha_{\perp}}

\makeatletter
\def\@dotsep{4.5}
\makeatother
\usepackage{times}
\usepackage{epsfig}
\usepackage{graphicx}
\usepackage{dcolumn}
\usepackage{bm}
\usepackage{xspace}

\begin{document}

\title{Femtosecond study of the effects of ions on the reorientation dynamics of water}

\author{Sietse T.\ van der Post}
\email{post@amolf.nl}
\author{Stefan Scheidelaar}
\author{Huib J.\ Bakker}
\affiliation{FOM Institute AMOLF, Science Park 104, Amsterdam, The
Netherlands }

\date{\today}
\begin{abstract}
We study the effects of ions on the reorientation dynamics of liquid
water with polarization-resolved femtosecond mid-infared
spectroscopy. We probe the anisotropy of the excitation of the
\mbox{O--D} stretch vibration of HDO molecules in solutions of NaCl,
NaI and N(alkyl)$_{4}$Br (tetra-alkylammoniumbromide) salts in 8\%
HDO:H$_{2}$O. We find that the reorientation \mbox{O--D} groups of
HDO molecules hydrating the Cl$^{-}$ and I$^{-}$ anions occurs on
two different time scales with time constants of 2$\pm$0.3 ps and
9$\pm$2 ps. The fast component is due to a wobbling motion of the
\mbox{O--D} group that keeps the hydrogen bond with the halogenic
anion intact. For solutions of N(alkyl)$_{4}$Br salts we observe a
very strong slowing down of the reorientation of water that is
associated with the hydration of the hydrophobic alkyl groups of the
N(alkyl)$_{4}$$^{+}$ ions.
\end{abstract}
\maketitle
\newpage
\section*{Introduction}
Water plays an essential role in the determination of the spatial
structure of (bio)molecules and (bio)molecular ions. Examples of
this role of water are the hydrophobic collapse in the folding of
proteins and the self-organized formation of cell membranes
\cite{lee04}. However, the interactions between water and a solute
not only determine the properties of the solute, these interactions
also change the structure and dynamics of water itself.

Water molecules close to a solute or interface tend to show slower
orientational dynamics than in bulk liquid water. For hydrophylic
solutes and surfaces the slowing down can be explained from the
specific interactions between the water molecules and the
solute/surface. These interactions can take the form of a
hydrogen-bond or can be purely electrostatic.
The electric fields exerted by ionic charges lead to a strong
ordering of the surrounding water. In earlier times it was thought
that the effects of ions on water structure were quite long range
and would extend well beyond the first hydration layer \cite{cox34}.
Some ions were thus referred to as structure makers, others as
structure breakers. However, recent experimental and theoretical
work showed that for most ions the effect on the structure and
dynamics of water is limited to only one or two hydration layers
\cite{omta03,smith07,mancinelli07,schmidt09,lin09}. Only for
specific combinations of ions, like Mg$^{2+}$ and SO$_{4}$$^{2-}$,
it was found that the water dynamics can be impeded over relatively
long ranges \cite{tielrooij10}.

Hydrophobic surfaces also influence the structure and dynamics of
water due to the fact that hydrophobic molecular groups take up
space in the hydrogen-bond network of the water liquid. This
occupation of space has profound effects on the free energy of the
solute and its hydration shell. These effects depend on the size of
the hydrophobe \cite{chandler05}. Small hydrophobes ($<$1 nm) can be
accommodated in the hydrogen-bond network of water without
significantly breaking of hydrogen bonds, similar as one may create
a hole in loosely woven fabric by distorting the pattern of the
threads without needing to break them \cite{muller90}. The water
molecules form tangential hydrogen bonds to each other along the
hydrophobic surface, and the water density near this surface is
larger than in the bulk \cite{chandler05}. For large hydrophobic
structures ($>$1 nm), the water hydrogen-bond network is truncated
near the hydrophobic surface. In this case, the water density is
lower near the surface than in the bulk. If two hydrophobes merge,
the solvation free energy is reduced, which forms the driving force
of the hydrophobic collapse. The occupation of space in the
hydrogen-bond network of water also explains the slowing down of the
dynamics of nearby water molecule. The reorientation of a water
molecule in bulk water involves the collective repositioning of
several water molecules, and these collective dynamics are hindered
by the presence of the hydrophobe
\cite{rezusjpca08,petersen09,laage09}.

In this paper we study the effects of different ions on the
orientational mobility of water molecules. We probe these dynamics
with polarization-resolved femtosecond mid-infrared spectroscopy. We
study the combined effects of electrostatic and hydrophobic
interactions on the dynamics of water by measuring the dynamics of
water in the hydration shells of different N(alkyl)$_{4}$$^{+}$
(tetraalkylammonium) ions.

\section*{Experiment}
We study the dynamics of water molecules in salt solutions by
monitoring the reorientation of HDO molecules in isotopically
diluted water (4\% D$_2$O in H$_2$O = 8\% HDO:H$_{2}$O). These
orientational dynamics are measured by probing the anisotropy
dynamics of excited \mbox{O--D} vibrations with
polarization-resolved femtosecond mid-infrared (fs-IR) pump-probe
spectroscopy. This technique requires intense femtosecond light
pulses at a wavelength of 4 $\mu$m (2500 cm$^{-1}$). These pulses
are obtained via parametric amplification processes that are pumped
by the pulses of a regeneratively amplified Ti:sapphire laser system
(Spectraphysics Hurricane).

The Ti:sapphire laser system generates pulses at a wavelength of
800~nm with a pulse energy of $\sim$900 $\mu$J at a repetition rate
of 1~kHz. The output of the laser is split into two parts. The first
part is used to pump a white-light seeded Optical Parametric
Amplifier (OPA) based on a $\beta$-bariumborate (BBO) crystal
(Spectra Physics). The BBO crystal is angle tuned to generate light
at 1.3 $\mu$ (signal) and 2 $\mu$m (idler). The idler is frequency
doubled in a second BBO crystal to a wavelength of 1 $\mu$m. This 1
$\mu$m pulse is used as a seed in a parametric amplification process
in a potassium niobate crystal (KNB) crystal that is pumped with the
remaining part of the 800~nm beam. In this latter process pulses are
generated with a central wavelength of 4 $\mu$m, a pulse energy of 5
$\mu$J, and a pulse duration of 120 femtoseconds.

The 4 $\mu$m pulses are sent onto a wedged CaF$_{2}$ plate. The
transmitted part ($\sim$90 \%) serves as the pump in the
polarization-resolved pump-probe experiment. The reflection from the
front side is sent into a variable delay stage with a resolution of
3~fs. This fraction forms the probe. With the delay stage we vary
the time delay $t$ between pump and probe. The reflection from the
back side of the wedged CaF$_{2}$ plate serves as a reference. The
pump is transmitted through a $\lambda$/2 plate to rotate its
polarization at 45 degrees with respect to the probe. The pump,
probe and reference are all focussed in the sample using a
gold-coated parabolic mirror, but only the pump and the probe are in
spatial overlap. After the sample, the probe and the reference are
dispersed with an Oriel monochromator and detected with the two
lines of an Infrared Associates 2$\times$32 MCT
(mercury-cadmium-telluride) detector array. The measurement of the
reference thus allows for a frequency-resolved correction for
shot-to-shot fluctuations of the probe-pulse energy.

The pump promotes population from the equilibrium ground-state $v=0$
to the first excited state $v=1$. This excitation leads to a
bleaching signal and stimulated emission at frequencies matching the
$v=0\rightarrow 1$ transition, and to an induced absorption at
frequencies matching the $v=1\rightarrow 2$ transition. The pump
pulse will preferentially excite \mbox{O--D} groups of which the
transition dipole moment is oriented parallel to the polarization of
the pump pulse. Hence, the excitation will be anisotropic. After the
sample, the polarization components of the probe parallel and
perpendicular to the polarization of the pump are selected with a
polarizer mounted on a rotation stage. Thereby we obtain the
pump-induced transient absorption changes for both polarization
directions: $\apar(\omega,t)$ and $\aper(\omega,t)$. These
absorption changes are used to determine the anisotropy of the
\mbox{O--D} excitation:
\begin{equation}\label{eq:aniOld}
R(\omega,t) = \frac{\apar(\omega,t)-\aper(\omega,t)}{\apar(\omega,t)+2\aper(\omega,t)},
\end{equation}
with $R(\omega,t)$ the anisotropy of the excitation that depends on
the probe frequency $\omega$ and the delay time $t$ between pump and
probe. The denominator of the above expression represents the
so-called isotropic signal and serves to divide out the decay of
$\apar(\omega,t)$ and $\aper(\omega,t)$ due to the vibrational
relaxation of the \mbox{O--D} excitation. As a result, $R(\omega,t)$
only reflects the orientational dynamics of excited \mbox{O--D}
vibrations. In the case of isotopically diluted samples, the
dynamics of $R(\omega,t)$ only represents the molecular
reorientation of the \mbox{O--D} groups of the HDO molecules.

We use the above described technique to study the anisotropy
dynamics of HDO molecules in solutions of NaCl, NaI and different
N(alkyl)$_{4}$Br salts in 8\% HDO:H$_{2}$O. All salts were of
$\geq$99.5\% purity and were purchased from Sigma Aldrich. D$_2$O
was also purchased from Sigma Aldrich and has a purity of
$\geq$99.99\%. The sample cell consisted of two 4 mm calcium
fluoride windows which were pressed against each other with a teflon
spacer in between. To keep the infrared transmission around 10\% we
used spacers of 25 $\mu$m for the samples with low salt
concentrations and 50 $\mu$m for the samples with high salt
concentrations. All concentrations in this paper are denoted in
moles of solute per kilogram of solvent (molality).

\section*{Results and Discussion}

\begin{figure} [h!!!!!]
   \includegraphics [scale=0.4]{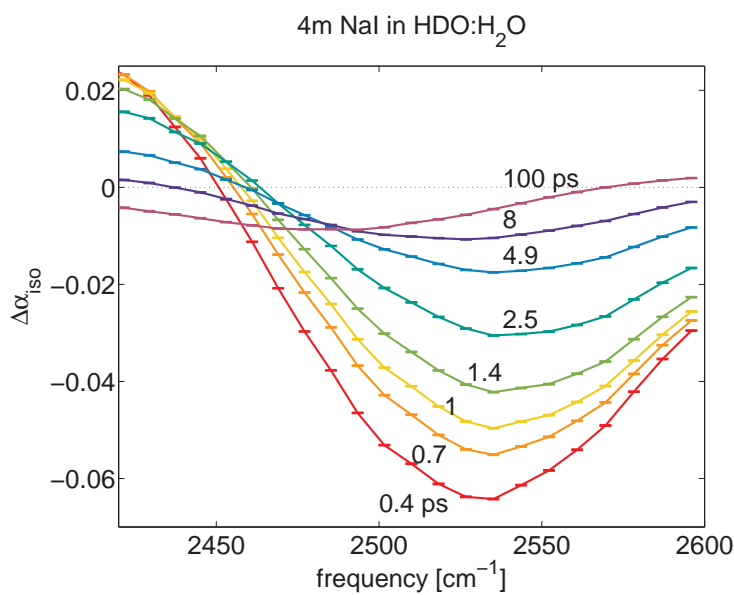}
   \caption{Transient isotropic spectra at different delay times after excitation of the \mbox{O--D} stretch vibration of HDO molecules
   for a solution of 4 m NaI in 8\% HDO:H$_{2}$O.}
   \label{transientspectra}
\end{figure}

Figure~\ref{transientspectra} shows transient absorption spectra of
the \mbox{O--D} stretch vibration of HDO molecules at different
delay times $t$ measured for a solution of 4 m NaI in 8\%
HDO:H$_{2}$O. These transient spectra are difference spectra between
the absorption spectrum measured at a particular delay time after
the excitation, and the original ground state absorption spectrum
(no excitation pulse). At early delays the transient spectra are
characterized by a negative (bleach) contribution at high
frequencies ($>\sim$2450~cm$^{-1}$) representing the bleached
absorption and stimulated emission of the $v=0\rightarrow$1
transition, and a positive signal at lower frequencies
($<\sim$2450~cm$^{-1}$), representing the induced $v=0\rightarrow$1
absorption. The bleaching signal and the induced absorption both
decay with a time constant of $\sim$2 ps.

After $\sim$10~ps the transient spectrum does not change any further
with increasing delay time. After this time the excited \mbox{O--D}
vibrations have completely relaxed and the transient spectrum
represents the spectral response resulting from the heating of the
sample by a few Kelvin. An increase in temperature induces a
decrease in the cross-section and a small blueshift of the
\mbox{O--D} stretch vibrations. The difference spectrum of the
absorption spectrum at long delay times and the original absorption
spectrum thus shows a persistent bleaching in the main part of the
\mbox{O--D} stretch absorption spectrum and a small induced
absorption in the blue wing of this spectrum.

We find that the vibrational relaxation of the \mbox{O--D}
vibrations does not immediately lead to a rise in temperature. A
similar delayed rise of the thermal response was observed in a study
of the vibrational dynamics of neat 8\% HDO:H$_2$O \cite{rezus05}.
This delayed thermal response can be due to the transient population
of a particular non-thermal state in the vibrational relaxation of
the \mbox{O--D} stretch vibration. Such a non-thermal state may
involve the \mbox{H--O--D} bending mode and/or the librational
modes. However, the intermediate state was observed to have no
spectral signature \cite{rezus05}, which means that its absorption
spectrum is identical to that of the \mbox{O--D} stretch vibration
before the excitation by the pump. Therefore, the delayed rise of
the thermal response is likely not due to the transient excitation
of a specific mode like the \mbox{H--O--D} bending vibration,
because such an excitation would lead to an anharmonic frequency
shift of the absorption spectrum of the \mbox{O--D} stretch
vibration. Instead, the delayed rise of the thermal spectrum is
likely due to the relatively slow adaptation of the coordinates of
low-energy degrees of freedom (hydrogen-bond bend and stretch) to
the higher energy content that results from the relaxation of the
\mbox{O--D} stretch vibration. The relaxation of the \mbox{O--D}
stretch vibration leads to a rapid increase of the energy content of
the lower-energy degrees of freedom, and these coordinates need some
time to evolve to the new equilibrium positions corresponding to
this higher energy content.

In previous work it was found that the vibrational relaxation
dynamics of salt solutions containing Cl$^{-}$, Br$^{-}$, or I$^{-}$
ions consists of two distinct components \cite{kropman01b,park07}.
One component is formed by \mbox{O--D} groups that are hydrogen
bonded to the oxygen atom of another water molecule. This component
comprises both the bulk water molecules and the water molecules in
the hydration shells of the cations. These \mbox{O--D}$\cdots$O
groups absorb at about the same frequency as the \mbox{O--D} groups
in neat HDO:H$_{2}$O and have a vibrational lifetime $T_{1}$ of
1.8$\pm$0.2 ps. The other component is formed by \mbox{O--D} groups
that are hydrogen bonded to the halogenic anion A$^{-}$(=Cl$^{-}$,
Br$^{-}$ or I$^{-}$). The absorption spectrum of these
\mbox{O--D}$\cdots$A$^{-}$ oscillators is blueshifted with respect
to the absorption spectrum of the \mbox{O--D}$\cdots$O component.
This blueshift results from the fact that the surface charge density
decreases with increasing anion size, which leads to a weakening of
the \mbox{O--D}$\cdots$A$^-$ hydrogen bond \cite{bergstroem91}. The
excited \mbox{O--D}$\cdots$A$^{-}$ vibrations have a vibrational
lifetime of 3-6 ps \cite{kropman01b,park07}, substantially longer
than the excited \mbox{O--D}$\cdots$O oscillators.

\begin{figure} [h!!!!!]
   \includegraphics [scale=1]{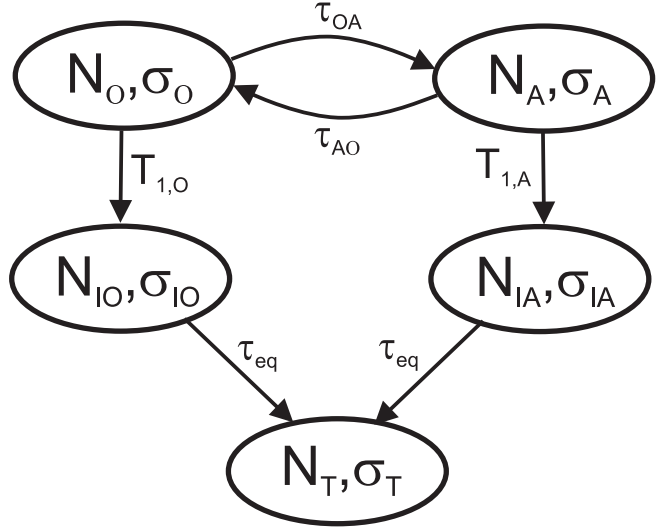}
   \caption{Schematic representation of the kinetic model used for fitting the vibrational decay processes.
   The populations $N_{O}$ and $N_{A}$ represent the excitation of the OD$\cdots$O and OD$\cdots$A$^-$ oscillators, respectively.
   These populations decay with respective relaxation time constants $T_{1,O}$ and $T_{1,A}$, and exchange with rate constant $\tau_{OA}$ and $\tau_{AO}$.
    The difference spectra $\sigma_{O}$ and $\sigma_{A}$ comprise the 1$\rightarrow$2 excited-state absorption, the 1$\rightarrow$0
    stimulated emission and the bleaching of the 0$\rightarrow$1 ground-state absorption. The presence of the latter contribution in
    $\sigma_{O}$ and $\sigma_{A}$ implies that the exchange dynamics of the ground states of the OD$\cdots$O and OD$\cdots$A$^-$
    oscillators are contained in the exchange of $N_{O}$ and $N_{A}$.}
   \label{model}
\end{figure}

We describe the vibrational relaxation with a model in which there
are two different excited states, corresponding to the $v=1$ states
of the \mbox{O--D}$\cdots$O and \mbox{O--D}$\cdots$A$^{-}$
oscillators. These excited states relax with vibrational relaxation
time constants $T_{1,O}$ and $T_{1,A}$ to their associated
intermediate states. These intermediate states then relax to the
final heated ground state with a time constant $\tau_{\rm eq}$. Each
of the states has a specific associated difference spectrum with
respect to the ground state. The shapes of the difference spectra
are fitted to the measured transient spectra at all different delay
times. The fitted difference spectra do not change in shape as a
function of delay time, they only change in amplitude following the
relaxation dynamics of the kinetic model that is schematically
depicted in Fig.\ \ref{model}.

In the kinetic modeling we included the effects of exchange between
the anion hydration shell and the bulk. Recently this exchange was
studied for aqueous solutions of NaBF$_{4}$ \cite{moilanen09b} and
NaClO$_{4}$ \cite{ji10,gaffney11}. The time constant of the
switching from the hydration shell of the anion to the bulk was
found to be $\sim$7 ps for BF$_{4}$$^{-}$ and $\sim$9 ps for
ClO$_{4}$$^{-}$. Unfortunately, we cannot measure this time constant
for Cl$^{-}$ and I$^{-}$, because the \mbox{O--D}$\cdots$A$^{-}$ and
the \mbox{O--D}$\cdots$O absorption bands are too strongly
overlapping for these salt solutions. For BF$_{4}$$^{-}$ and
ClO$_{4}$$^{-}$ the \mbox{O--D}$\cdots$A$^{-}$ and the
\mbox{O--D}$\cdots$O absorption bands are quite well separated, thus
allowing the measurement of the exchange rate between these
components. In view of the fact that the hydrogen bonds between
water and the halogenic anions are stronger than those between water
and BF$_{4}$$^{-}$/ClO$_{4}$$^{-}$, we expect the switching time for
rotation out of the hydration shell of Cl$^{-}$ and I$^{-}$ to the
bulk to be at least as large as observed for BF$_{4}$$^{-}$ and
ClO$_{4}$$^{-}$, i.e.\ $\tau_{\rm AO}\leq$9 ps. In the modeling we
use a value of $\tau_{\rm AO}$ of 9 ps. The value of $\tau_{\rm OA}$
of the reverse process was chosen such that the number density of
\mbox{O--D}$\cdots$A$^{-}$ oscillators is conserved. Hence,
$\tau_{\rm AO}/\tau_{\rm OA} = N_{{\rm O-D}\cdots {\rm O}}/N_{{\rm
O-D}\cdots {\rm A}^{-}}$, meaning that $\tau_{\rm OA}$ will decrease
with increasing salt concentration.

\begin{figure} [h!!!!!]
   \includegraphics [scale=0.8]{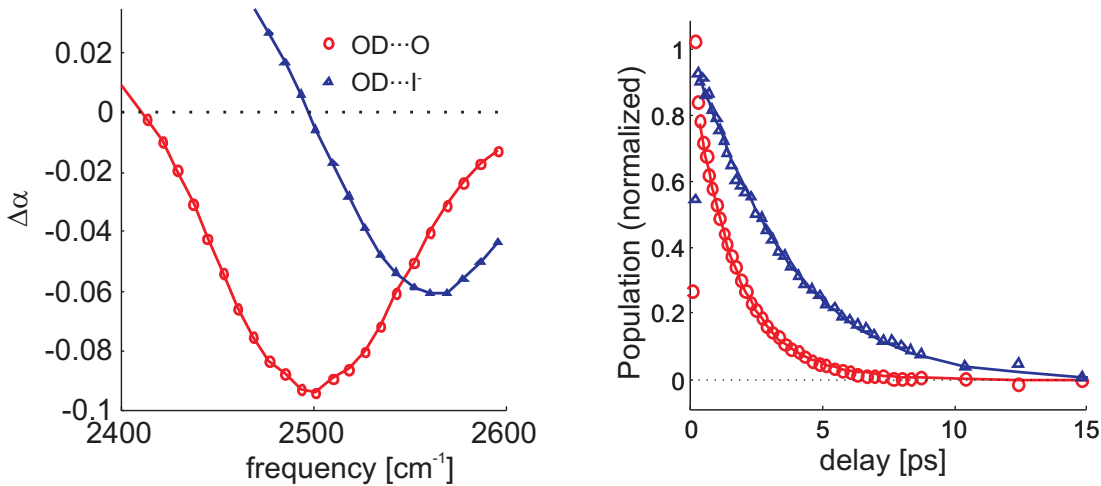}
   \caption{Left panel: Spectral components obtained from fitting the spectra measured for a 4 m NaI solution
   presented in Fig.\ 1. Right panel: Delay time dependence of the populations of the spectral components. At any
   given delay-time the fitted total transient spectrum is the weighted sum of the spectra shown in
   the left panel, the weight factors given by the population values in the right panel.}
   \label{model_decomposition}
\end{figure}

Fig.\ \ref{model_decomposition} presents the results of a fit of the
above described kinetic model to the data shown in Fig.\
\ref{transientspectra}. The left panel of Fig.\
\ref{model_decomposition} shows the difference spectra associated
with the excited \mbox{O--D}$\cdots$O vibration and the excited
\mbox{O--D}$\cdots$I$^{-}$ vibration, resulting from the fit. For
the two excited states the difference spectra consist of a bleaching
of the fundamental 0$\rightarrow$1 absorption and an induced
1$\rightarrow$2 excited-state absorption. The right panel of figure
\ref{model_decomposition} presents the associated population
dynamics. It is clearly seen that the excited \mbox{O--D}$\cdots$O
and \mbox{O--D}$\cdots$I$^{-}$ states show very different
vibrational relaxation time constants $T_{1}$ of 1.8$\pm$0.2 and
4.6$\pm0.4$ ps, respectively. The equilibration time constant
$\tau_{\rm eq}$ is 1$\pm$0.2 ps.

\begin{figure} [h!!!!!]
   \begin{center}$
   \begin{array}{cccc}
   {\mbox{\bf a}}
   \includegraphics [scale=.3]{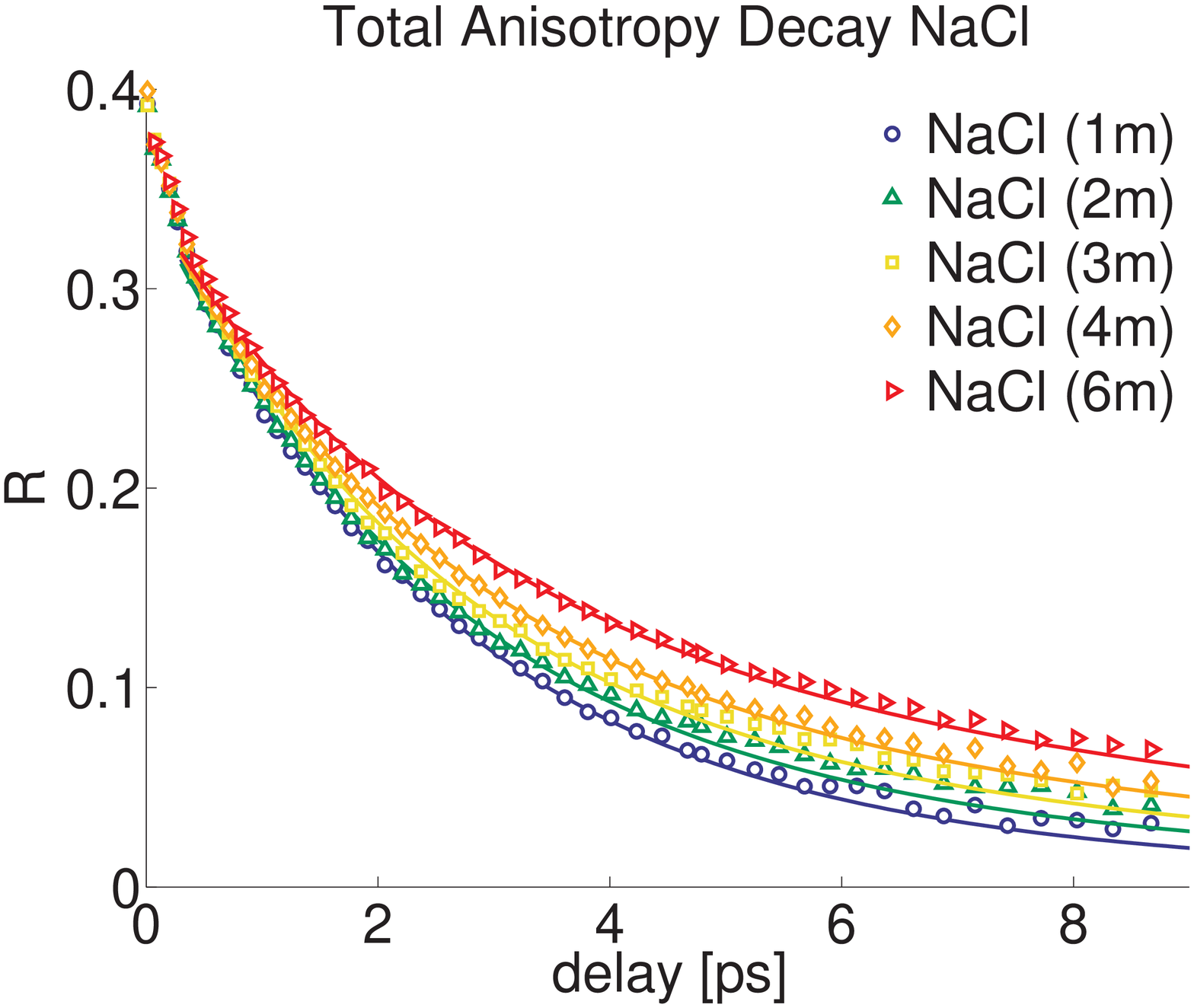}
   {\mbox{\bf b}}
   \includegraphics [scale=.3]{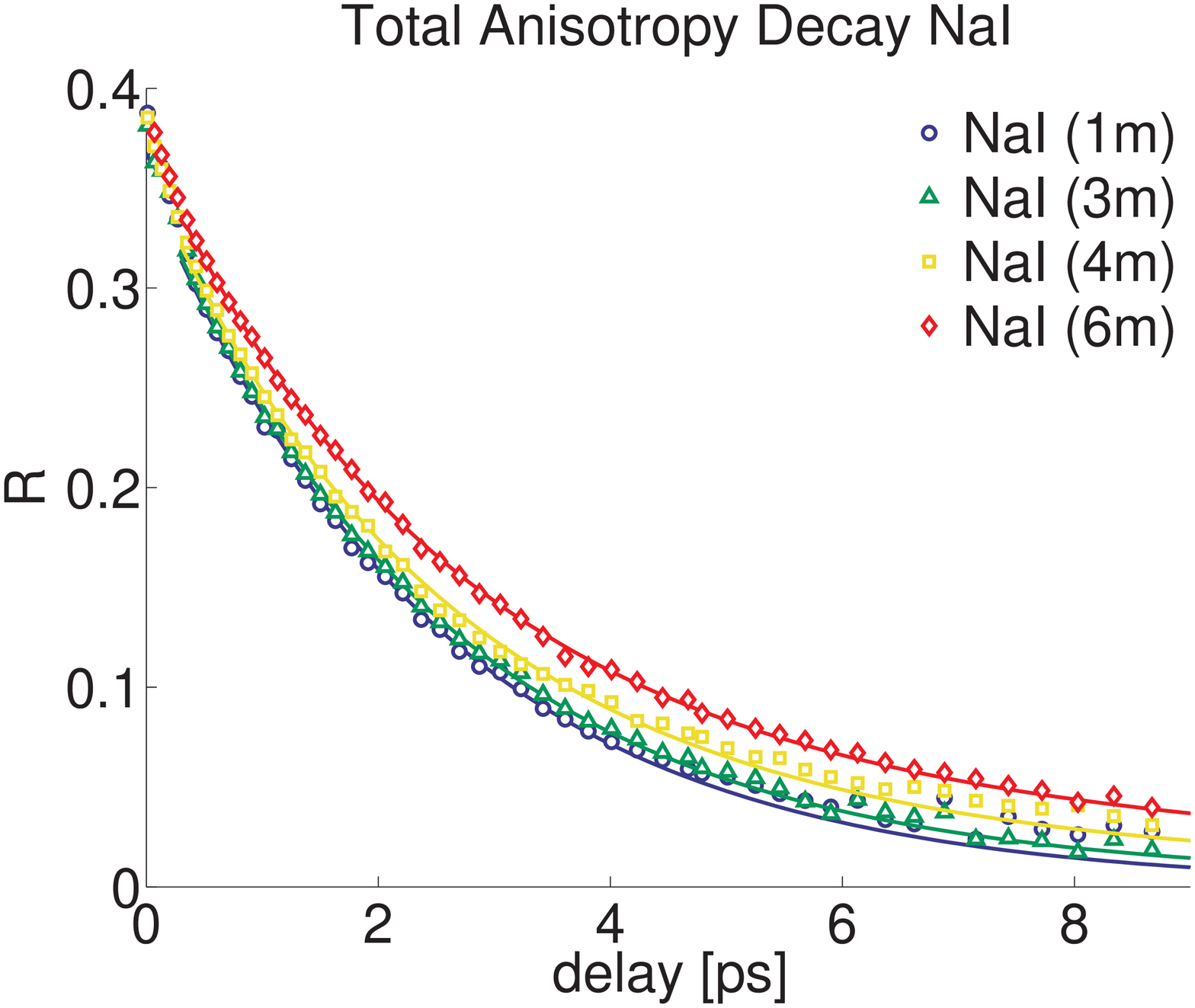}
   \end{array}$
   \end{center}
   \caption{(\textbf{a}) Anisotropy of the \mbox{O--D} stretch vibration of HDO
   as a function of delay time for solutions of different concentrations
   of NaCl in 8\% HDO:H$_{2}$O. The solid lines result from a bi-exponential fit to the data.
   (\textbf{b}) As Fig.\ (\textbf{a}) for solutions containing different concentrations
    of NaI. }
   \label{AnisotropyNaClNaI}
\end{figure}

To obtain the anisotropy dynamics of the excited \mbox{O--D} stretch
vibrations we need to subtract the delay-time dependent contribution
of the heated ground state from the measured transient spectra. The
spectral response associated with the final heated ground state is
observed to be isotropic, meaning that we have to subtract the same
amplitude for this spectrum from the measured $\apar(\omega,t)$ and
$\aper(\omega,t)$. The resulting $\apar(\omega,t)$ and
$\aper(\omega,t)$ are then used to construct the delay-time
dependent anisotropy of the \mbox{O--D} stretch vibrations using
equation (\ref{eq:aniOld}). In Fig.\ \ref{AnisotropyNaClNaI} we show
the thus obtained anisotropy dynamics measured for solutions of
different concentrations of NaCl and NaI. For both types of salt
solutions it is observed that the anisotropy dynamics become slower
with increasing salt concentration. For the NaCl solutions this
effect is more pronounced than for the NaI solutions.

The decomposition of the measured transient spectra in
\mbox{O--D}$\cdots$O and \mbox{O--D}$\cdots$A$^{-}$ components also
enables the determination of the anisotropy dynamics for each of
these components separately. To this purpose, we construct the
difference spectrum $D(\omega,t)$ between $\apar(\omega,t)$ and
$\aper(\omega,t)$. It can easily be shown that $D(\omega,t)$ is
proportional to the weighted sum of the anisotropy values $R_{i}$ of
the different spectral components, the weighing factor of each
component given by the product of the spectral amplitude
$\sigma_{i}(\omega)$ at frequency $\omega$ and the population
$N_{i}(t)$ of the component at delay time $t$:
\begin{eqnarray}\label{eq:diff}
D(\omega,t) = \frac{\apar - \aper}{3} =
\displaystyle\sum\limits_{i=1}^n R_i(t)N_i(t)\sigma_i(\omega)
\end{eqnarray}
The spectra $\sigma_{i}(\omega)$ and the populations $N_{i}(t)$ are
obtained from the fit of the isotropic data. Using these spectra and
populations we can determine the time-dependent anisotropy
$R_{i}(t)$ of each of the contributing spectral components. This
approach assumes that the anisotropy dynamics is the same at all
frequencies within the spectrum $\sigma_{i}(\omega)$ of each
component. In this analysis we only need to consider the
\mbox{O--D}$\cdots$O and \mbox{O--D}$\cdots$A$^{-}$ spectral
components, because the intermediate state has no associated
difference spectrum ($\sigma_{\rm IO}(\omega)=\sigma_{\rm
IA}(\omega)=0$), and the final heated ground state has zero
anisotropy ($R_{\rm T}(t)=0$).

\begin{figure} [h!!!!!]
   \includegraphics [scale=.4]{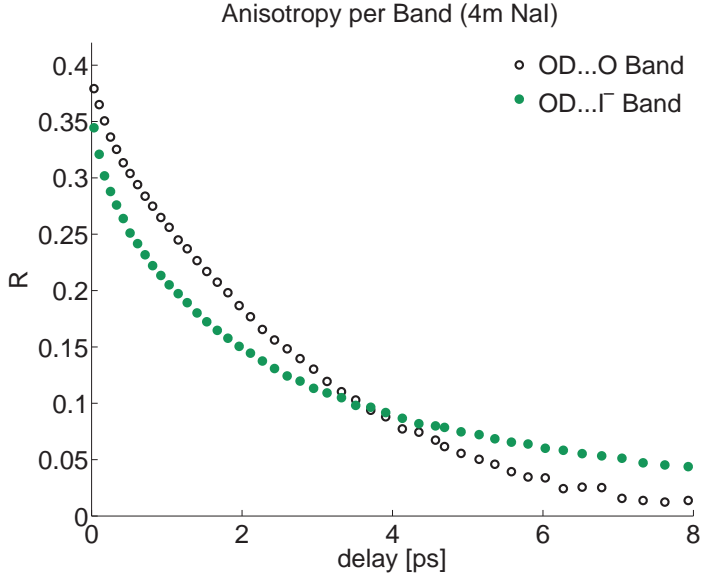}
   \caption{Anisotropy dynamics of the spectral components associated with \mbox{O--D}$\cdots$O oscillators (black open circles)
   and \mbox{O--D}$\cdots$I$^{-}$ oscillators (green solid circles). The anisotropy decay of the \mbox{O--D}$\cdots$O oscillators
   can be fitted well to a single exponential decay with a time constant of 2.5 ps, and is thus very similar to the
   dynamics observed for neat HDO:H$_{2}$O. The anisotropy decay of the \mbox{O--D}$\cdots$I$^{-}$ oscillators can be
   fitted well to a bi-exponential function with time constants of 2 and 9 ps.}
   \label{anisotropy_decomposition}
\end{figure}

In Fig.\ \ref{anisotropy_decomposition} the resulting $R_{O}(t)$ of
the \mbox{O--D}$\cdots$O component and $R_{A}(t)$ of the
\mbox{O--D}$\cdots$A$^{-}$ component are shown for a 4 m NaI
solution. The dynamics of $R_{O}(t)$ can be fitted well with a
single exponential function with a time constant of $\sim$2.5 ps,
which means that the anisotropy dynamics of the \mbox{O--D}$\cdots$O
oscillators is very similar to the dynamics observed for neat
HDO:H$_2$O \cite{rezus05}. The dynamics of $R_{A}(t)$ is seen to
deviate strongly from a single exponential function. The dynamics of
$R_{A}(t)$ follow a bi-exponential function with time constants of
2$\pm$0.3 and 9$\pm$2 ps.

\begin{figure} [h!!!!!]
   \includegraphics [scale=0.8]{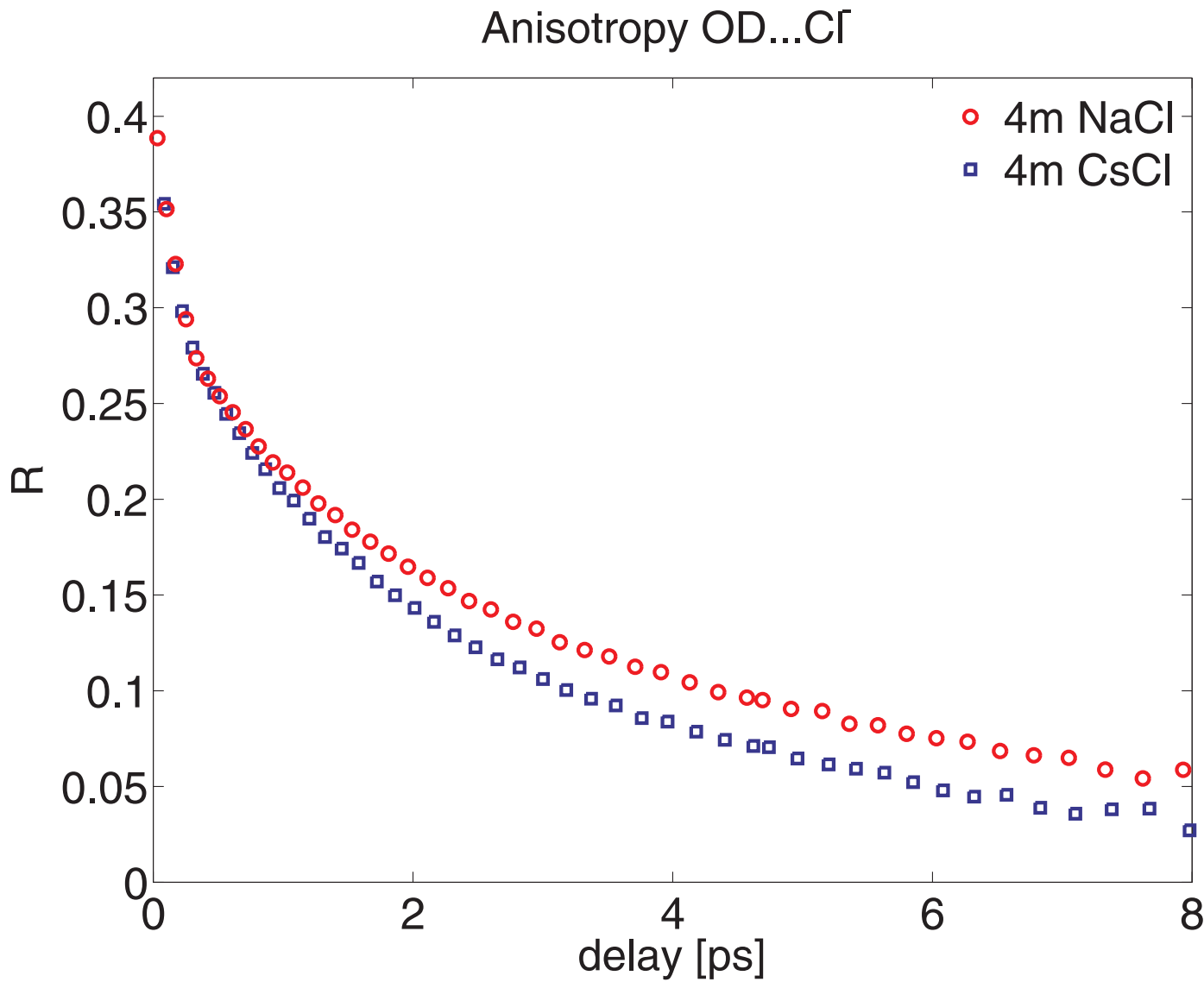}
   \caption{Anisotropy dynamics of the spectral components associated with \mbox{O--D}$\cdots$Cl$^{-}$ oscillators
   (red circles) and \mbox{O--D}$\cdots$I$^{-}$ oscillators (blue circles). Both anisotropy decays can be fit
   well to a bi-exponential function with time constants of 2 and 9 ps, but with a larger amplitude for the
   fast (2 ps) component for the \mbox{O--D}$\cdots$Cl$^{-}$ component than for the \mbox{O--D}$\cdots$I$^{-}$ component.}
   \label{comparison_nai_nacl}
\end{figure}

In Fig.\ \ref{comparison_nai_nacl} we compare the $R_{A}(t)$ curves
of solutions of 4 m NaI and 4 m NaCl. It is seen that the dynamics
are quite similar, both anisotropy decay curves can be fitted well
with a bi-exponential function with time constants of 2 and 9 ps.
However, the amplitude of the fast component is larger for the 4 m
NaCl solution than for the 4 m NaI solution.

The decay of the anisotropy of the \mbox{O--D}$\cdots$A$^{-}$
oscillators is not due to the rotation of \mbox{O--D} oscillators
out of the hydration shell. These oscillators exchange their
\mbox{O--D}$\cdots$A$^{-}$ hydrogen bond for an \mbox{O--D}$\cdots$O
hydrogen bond and thus no longer contribute to the
\mbox{O--D}$\cdots$A$^{-}$ spectral component. Therefore, the
rotation out of the hydration shell will only lead to a decay of the
amplitude of the \mbox{O--D}$\cdots$A$^{-}$ spectral component, but
not of the anisotropy of this component. However, there is an equal
number of \mbox{O--D} groups that were excited as
\mbox{O--D}$\cdots$O oscillators and have changed their orientation
to become \mbox{O--D}$\cdots$A$^{-}$ oscillators. These oscillators
will contribute to the measured anisotropy of the
\mbox{O--D}$\cdots$A$^{-}$ spectral component. In fact it can be
expected that these oscillators will decrease the anisotropy of the
\mbox{O--D}$\cdots$A$^{-}$ spectral component because the angle of
rotation in the exchange between the hydration shell and the bulk is
$\sim$55 degrees \cite{laagescience06,laage07,laage08}. The time
constants of the exchange are of the order of 10 ps, and thus the
switching from the bulk to the hydration shell cannot form the
origin of the fast decay of the anisotropy of the
\mbox{O--D}$\cdots$Cl$^{-}$ and \mbox{O--D}$\cdots$I$^{-}$ spectral
components. This switching will however contribute to the slow
component of the anisotropy decay. Other effects that can be
responsible for the slow decay are the diffusion of the \mbox{O--D}
group over the surface of the ion, and the reorientation of the
complete hydration shell of the halogenic anion.

\begin{figure} [h!!!!!]
   \includegraphics [scale=.8]{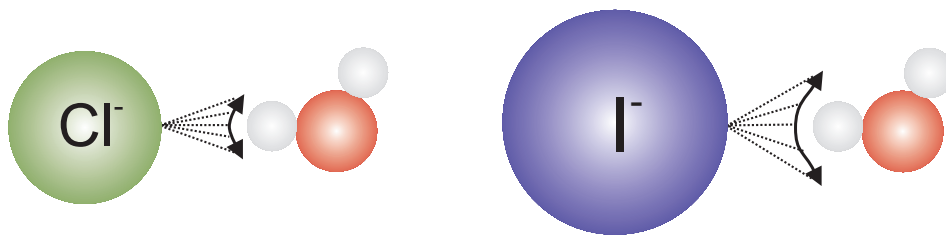}
   \caption{Schematic picture of the wobbling component of the \mbox{O--D}$\cdots$A$^{-}$ groups in the hydration shells
   of Cl$^{-}$ and I$^{-}$. The larger size of I$^{-}$ allows for a larger variation of the angle between \mbox{O--D} and
   the O$\cdots$A$^{-}$ coordinate without significantly weakening the \mbox{O--D}$\cdots$A$^{-}$ hydrogen bond.}
   \label{cartoon_nai_nacl}
\end{figure}

The fast component of the anisotropy decays shown in Fig.\
\ref{comparison_nai_nacl} is likely due to a wobbling motion of the
\mbox{O--D} group that keeps the \mbox{O--D}$\cdots$A$^{-}$ hydrogen
bond with the anion intact. The wobbling motion will lead to a
partial decay of the anisotropy, and the amplitude of this partial
decay is determined by the angular cone of the wobbling motion. The
I$^{-}$ ion is much larger than the Cl$^{-}$ ion and thus allows for
a larger angular spread of the \mbox{O--D} group while keeping the
hydrogen bond intact. The effect of the ion size on the angular cone
of the wobbling motion is illustrated in Fig.\
\ref{cartoon_nai_nacl}. The wobbling motion will be induced by the
translational and orientational motions of the H$_{2}$O molecules
that surround the hydration shell. Therefore, the time constant of
the wobbling motion can be expected to be similar to that of the
anisotropy decay of the \mbox{O--D}$\cdots$O component. It may be
even somewhat faster, as the wobbling of the \mbox{O--D} group in
the anion hydration shell does not require the completed
reorientation of an \mbox{O--D} group outside the hydration shell.
The time constant of the fast decay of the anisotropy of 2$\pm$0.3
ps agrees with this picture.

We thus find that the reorientation of the water molecules in the
hydration shells of Cl$^{-}$ and I$^{-}$ possess a significant
wobbling component with a time constant that is quite similar to
that of the reorientation in bulk liquid water, even at very high
concentrations of 4 m. This means that these weakly hydrating anions
do not show a strong effect on the dynamics of water. At first
sight, the cations show even less effect, the dynamics of the
\mbox{O--D}$\cdots$O component includes the water molecules that
hydrate the cations, but even for solutions containing high
concentrations of K$^{+}$ and Na$^{+}$, the dynamics of the
\mbox{O--D}$\cdots$O component are very similar to the dynamics
observed for neat HDO:H$_{2}$O. This observation suggests that
K$^{+}$ and Na$^{+}$ hardly modify the reorientation dynamics of the
water molecules in their hydration shells. However, at this point it
should be noted that in the present experiments we only study the
orientational motions of the hydroxyl groups of the water molecules.
Cations do show a strong effect on the reorientation of the dipole
moments of the water molecules, as has been observed with dielectric
relaxation studies \cite{buchner99,turton08}. In these studies it
was found that cations like Li$^{+}$, Na$^{+}$ and Mg$^{2+}$ fix the
orientation of the dipole moment of a number of water molecules in
their first hydration shell. The present work demonstrates that the
hydroxyl groups of these water molecules remain quite mobile,
meaning that these groups must reorient in a propeller-like motion
around the dipole that is fixed by the electric field of the cation
\cite{tielrooij11}.

\begin{figure} [h!!!!!]
   \includegraphics [scale=0.9]{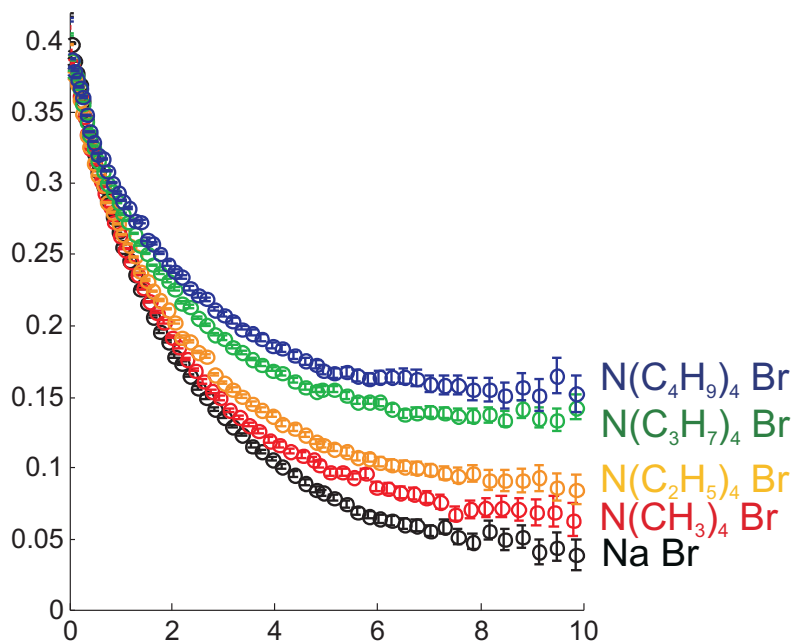}
   \caption{Anisotropy of the \mbox{O--D} stretch vibration of HDO
   as a function of delay time for 1 m solutions of different tetra-alkylammoniumbromide salts in 8\% HDO:H$_{2}$O.}
   \label{ani_taabrsalts}
\end{figure}

In previous studies we found that hydrophobic molecular groups
strongly affect the reorientation dynamics of water
\cite{rezus07,rezus08,tielrooij10b}. The addition of amphiphilic
molecules to water like trimethylaminoxide (TMAO) and
tetramethylurea (TMU) was observed to lead to a strong slowing down
of the reorientation dynamics of a number of water molecules. These
molecules could be dissolved in water up to high concentrations due
to the presence of a polar group in the molecule. At low
concentrations, the fraction of slow water scales with the
concentration and with the size of the hydrophobic part of the
solute. At high concentrations, saturation effects occur which can
be very pronounced in case the solutes cluster, as is for instance
the case for TMU.

We investigate whether ions containing hydrophobic molecular groups
can have similar strong effects on the orientation dynamics of
water. In Fig.\ \ref{ani_taabrsalts} we show the anisotropy of the
\mbox{O--D} stretch vibration of HDO molecules for 1 m solutions of
different N(alkyl)$_{4}$Br (tetra-alkylammoniumbromide) salts in 8\%
HDO:H$_{2}$O. These anisotropy curves are obtained in a similar way
as the curves of Fig.\ \ref{AnisotropyNaClNaI}, meaning that the
anisotropy is constructed after the measured $\apar(\omega,t)$ and
$\aper(\omega,t)$ have been corrected for the contribution of the
ingrowing final heated ground state. For comparison also the
anisotropy curve of a 1 m solution of NaBr is shown. The anisotropy
curves shown in Fig.\ \ref{ani_taabrsalts} include both the
anisotropy dynamics of the \mbox{O--D} groups that hydrate the
anion, forming \mbox{O--D}$\cdots$Br$^{-}$ hydrogen bonds, and the
anisotropy dynamics of the \mbox{O--D} groups that are hydrogen
bonded to the oxygen atom of another water molecule, thus forming
\mbox{O--D}$\cdots$O hydrogen bonds.

The anisotropy dynamics observed for a 1 m solution of NaBr are very
similar to the anisotropy dynamics of neat HDO:H$_{2}$O, thus
confirming that Na$^{+}$ and a weakly hydrating anion like Br$^{-}$
do not have a strong effect on the reorientation dynamics of the
hydroxyl groups of water. In contrast, the N(alkyl)$_{4}$Br salts do
show a strong effect on the reorientation dynamics. Fig.\
\ref{ani_taabrsalts} shows that for these solutions there is a
fraction of water showing very slow dynamics with a time constant
$>$10 ps. As the Br$^{-}$ ion is not very effective in slowing down
the water reorientation, the slow water molecules are most likely
hydrating the N(alkyl)$_{4}$$^{+}$ cations. The fraction of slow
water clearly increases with increasing size of the alkyl groups of
the N(alkyl)$_{4}$$^{+}$ ion. Based on these observations we
conclude that the slow water molecules are hydrating these alkyl
groups. Hence, the effects observed of these hydrophobic groups on
the dynamics of water are quite similar to the effects we observed
before for uncharged amphiphilic solutes like TMAO and TMU
\cite{rezus07,rezus08,tielrooij10b}. The presence of a positive
charge on the N(alkyl)$_{4}$$^{+}$ ion apparently does not change
the effects of hydrophobic hydration on the dynamics of water.

Unfortunately, we could not analyze the dynamics of the
\mbox{O--D}$\cdots$O and the \mbox{O--D}$\cdots$Br$^{-}$ oscillators
separately as we did for the NaCl and NaI solutions, because such an
analysis requires the anisotropy dynamics to be uniform within each
of the components. For N(alkyl)$_{4}$Br solutions the anisotropy of
the \mbox{O--D}$\cdots$O and \mbox{O--D}$\cdots$Br$^{-}$ components
appears to be quite strongly frequency dependent. This frequency
dependence is likely due to the fact that the solutions containing
N(alkyl)$_{4}$$^{+}$ ion are strongly heterogeneous and show
relatively slow spectral diffusion dynamics, as indicated by their
high viscosity. As a result, the water molecules remain in different
environments for a long time and their different orientation
dynamics do not average out, as was the case for the much less
viscous NaCl and NaI solutions.

\begin{figure} [h!!!!!]
   \includegraphics [scale=0.8]{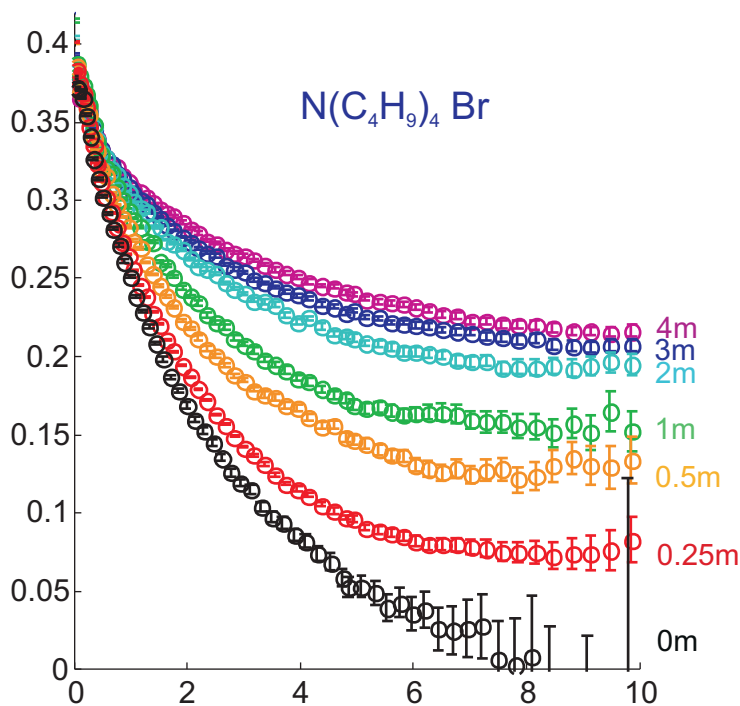}
   \caption{Anisotropy of the \mbox{O--D} stretch vibration of HDO
   as a function of delay time for solutions of different concentrations of tetrabutylammoniumbromide in 8\% HDO:H$_{2}$O.}
   \label{ani_tbabr}
\end{figure}

Fig.\ \ref{ani_tbabr} shows the anisotropy dynamics of the
\mbox{O--D} stretch vibration of HDO molecules measured for
solutions of N(C$_{4}$H$_{9}$)$_{4}$Br in 8\% HDO:H$_{2}$O of
different concentration. It is clearly seen that the fraction of
slow water strongly increases with the salt concentration. However,
there is a strong saturation effect: the fraction of slow water
strongly increases in the concentration interval between 0 and 1 m
and only shows a minor further increase in the interval from 2 to 4
m. This saturation of the fraction of slow water points at a
clustering of the N(C$_{4}$H$_{9}$)$_{4}$$^{+}$ ions. As the butyl
tails of the ions cluster, less of their surface will be exposed to
water, and this will reduce the amount of water being affected in
its reorientation dynamics. Apparently, the hydrophobic driving
force for clustering of the butyl groups of the
N(C$_{4}$H$_{9}$)$_{4}$$^{+}$ ions is stronger than the coulombic
repulsion of the positive charges of these ions.

\section*{Conclusions}
We studied the effects of ions on the reorientation dynamics of
water using polarization-resolved femtosecond mid-infrared
pump-probe spectroscopy. In these experiments we probe the
anisotropy dynamics of the \mbox{O--D} stretch vibration of HDO
molecules of solutions of salts dissolved in 8\% HDO:H$_{2}$O. For
solutions of NaCl and NaI we observe a moderate slowing down of the
anisotropy dynamics and thus of the reorientation of the water
molecules. We were able to distinguish the reorientation dynamics of
the \mbox{O--D} groups forming hydrogen bonds to other water
molecules from the dynamics of the \mbox{O--D} groups that form a
hydrogen bond to the halogenic anion.

The fraction of water molecules forming a hydrogen bond to another
water molecule includes the water molecules in the first hydration
shells of the Na$^{+}$ cations. For these water molecules we observe
surprisingly little effect of the interactions with the ions on the
reorientation dynamics. Even at high salt concentrations, the
\mbox{O--D} groups in the hydration shell of Na$^{+}$ show
reorientation dynamics that are quite similar to the dynamics
observed for neat liquid water, i.e.\ the orientational correlation
function shows a single exponential decay with a time constant of
$\sim$2.5 ps. However, previous dielectric relaxation studies showed
that Na$^{+}$ does slow down the reorientation of the dipole moments
of the water molecules in the first hydration shell
\cite{buchner99,turton08}. Hence, the hydroxyl groups of these water
molecules likely reorient in a propeller-like motion around the
water dipoles that are fixed by the electric field of Na$^{+}$.

The anisotropy of the \mbox{O--D} groups that are hydrogen bonded to
Cl$^{-}$ or I$^{-}$ decays on two different time scales. The first
time scale of $\sim$2 ps is associated with a wobbling motion of the
\mbox{O--D} group that keeps the hydrogen bond to the Cl$^{-}$ or
I$^{-}$ anion intact. This wobbling motion leads to a partial decay
of the anisotropy. The amplitude of this fast component is larger
for I$^{-}$ than for Cl$^{-}$, which can be explained from the fact
that the larger I$^{-}$ ion allows for a larger wobbling angle while
keeping the hydrogen bond intact, as illustrated in Fig.\
\ref{cartoon_nai_nacl}. The second time scale of $\sim$9 ps of the
anisotropy decay can result from the diffusion motion of the intact
\mbox{O--D} group over the surface of the ion, from the
reorientation of the complete ion hydration shell, and from the
switching of \mbox{O--D} groups into the hydration shell of the
anion. In previous studies only the $\sim$9 ps component was
observed, and from this observation it was concluded that the
solvation shell is rather rigid. In this study we could for the
first time separate the dynamics of the water molecules in the
hydration shell of chloride and iodide from that of the other water
molecules at all delay times. We thus find that the dynamics of the
water molecules in the hydration shell of chloride and iodide are
dominated by a large-amplitude wobbling motion. This is an important
result as it demonstrates the deformable character of these
hydration shells.

We observe a very strong slowing down of the water reorientation
dynamics for solutions of N(alkyl)$_{4}$Br
(tetra-alkylammoniumbromide) salts. For these solutions we observe a
fraction of slow water with a reorientation time constant $>$10 ps.
This fraction increases with the concentration of dissolved salt and
with the size of the alkyl groups of the N(alkyl)$_{4}$$^{+}$ ions.
These slow water molecules are thus likely hydrating the hydrophobic
parts of the N(alkyl)$_{4}$$^{+}$ ions. Hydrophobic molecular groups
thus have a very strong effect on the dynamics of water, even when
these hydrophobic groups are embedded in an ion. We thus find that
the effects of hydrophobic molecular groups on the dynamics of water
are much stronger than the electrostatic effects of ions. This is an
important result as the influence of hydrophobic and charged groups
on water plays a crucial role in the self-organizing dynamics of
biomolecular systems like proteins and membranes. For solutions of
N(C$_{4}$H$_{9}$)$_{4}$Br (tetra-butylammoniumbromide) we observe a
strong saturation effect of the fraction of slow water with
concentration, which indicates that these ions cluster, in spite of
the coulombic repulsion of the positive charges of the
N(C$_{4}$H$_{9}$)$_{4}$$^{+}$ ions.

\begin{acknowledgments}
This work is part of the research program of the ``Stichting voor
Fundamenteel Onderzoek der Materie (FOM)", which is financially
supported by the ``Nederlandse organisatie voor Wetenschappelijk
Onderzoek (NWO)".
\end{acknowledgments}

\end{document}